\documentclass[preprint,showpacs,preprintnumbers,amsmath,amssymb,floatfix]{revtex4}
\usepackage{graphicx}
\usepackage{dcolumn}
\usepackage{bm}
\usepackage{longtable}
\begin{document}

\title{ Size dependence of lattice constants of semiconductor nanocrystals}
\author{Roby Cherian, Priya Mahadevan}
\affiliation{ S.N. Bose National Centre for Basic Sciences, JD-Block, Sector III,Salt Lake, Kolkata-700098, India }
\date{\today}

\begin{abstract}
We have theoretically examined the size dependence of the 
equilibrium lattice constant of nanocrystals 
of Si, GaAs and CdSe. While deviations from the bulk lattice constant are 
as large as 1-2\% for unpassivated nanocrystals of Si, the deviations drop to
$\sim$ 0.3\%-0.4\% once the surfaces are passivated. Inspite of the fact that 
the average equilibrium bond-lengths are bulk-like, we find 
that the nearest-neighbor 
bond-length exhibits an unusual strain profile 
with bulk like bond-lengths in the 
core and shorter ($\sim$ 1\%) bonds at the surface.
\end{abstract}

\maketitle

The properties of semiconductor nanocrystals have received a lot of 
attention in the past two decades as a result of strong size dependence 
of various physical properties. The underlying crystal structure plays 
an important role in determining the electronic structure and the 
ensuing physical properties. However the basic issue of structure 
determination is difficult in the context of materials at the 
nanoscale. Even the most powerful techniques that one is familiar
with for the characterization of bulk crystals are found to fail in the case of nanocrystals 
\cite{billinge-science}. 
The difficulties arise because in the present case one does not 
have a homogeneous distribution of particles. Not only do the 
sizes vary over several percent depending on the synthesis procedure,
one also finds different shaped particles as a result of the 
growth techniques. Further most of the commonly used techniques provide 
an averaged lattice constant / bond-length. Hence several of the basic 
questions in the context of nanoparticles remain unanswered.
It is here that theoretical calculations play an important role. They 
can simulate the ideal situation and hence can be used to provide 
insight into the modifications in the lattice constant that take place 
as a function of nanoparticle size.

Nanocrystals consist of a sizeable number of atoms on the surface. In 
an ideal situation, all atoms at the surface would have broken 
coordination, while the rest would have bulk-like coordination. To make 
up for the lost coordination, stress develops at the surface and decays 
into the bulk. Naively as a result of this one would have longer / bulk-like 
bonds in the interior and shorter bonds at the surface.
The naive picture would be modified as the surface atoms are usually 
passivated by ligands and so the strain effects are not as strong as 
one would expect in the unpassivated case.

The experimental literature has conflicting reports of 
bulk-like nearest-neighbor bond-lengths \cite{heron}, 
small-intermediate strains resulting in modified lattice constants 
for the nanocrystals \cite{billinge-pdf} as compared with the 
bulk. However, the issue of the development and variation of strain is 
difficult to obtain experimentally. In this context theoretical 
models will be able to address the issue, as well as ascertain 
the role of the passivants.

We construct nanocrystals by cutting a spherical fragment of a
bulk crystal, which has the underlying geometry of the zinc-blende 
lattice (Si, GaAs and CdSe) / wurtzite lattice (CdSe) \cite{conflict}.
Now to define a spherical nanocrystal in this way we need to specify
the center of the sphere and the radius. In all our studies the nanocrystal is centered
on one atom and then the remaining atoms are generated by 
adding all atoms within the sphere of pre-defined radius and 
maintaining bulk-like geometry. Consequently the generated zinc-blende 
nanocystals will have a T$_d$ point group symmetry.
This prescription for constructing nanocrystals is similar to what has been 
used earlier in the literature \cite{gan-clusters}. It is believed to be valid for large 
cluster sizes as various experiments have found bulk-like coordination \cite{bulk-like}.
Table I provides the cluster sizes, defined in terms of the number of layers 
around the central atom, as well as the number of atoms of types A and B 
in the cluster with atom A at the center. 
As we use a plane wave implementation of density functional theory and 
are constrained to work with periodic systems, the electronic 
properties of the clusters are calculated by considering 
periodic clusters separated by $\sim$ 10 $\AA$ so that 
the interaction between the clusters is very small. 

The electronic structure of the clusters was 
calculated within plane wave
pseudopotential calculations using the implementation in VASP \cite{vasp}.
The semi core 3$d$ states on the Ga for the GaAs cluster calculations were
treated as a part of the core. For silicon we used the LDA approximation 
for the exchange while the GGA PW91 approximation \cite{ggapw91} to the 
exchange has been used for CdSe and GaAs and the calculations were performed at
Gamma point alone. A cut off energy of 250.0 eV was used for the plane wave basis
for Si and GaAs calculations while we used 274.3 eV for CdSe.

Initially the equilibrium 
bond-length is determined by minimizing the energy with respect to the 
lattice constant allowing for a uniform expansion or contraction of 
the volume. The equilibrium lattice constant was determined by 
fitting the energy as a function of volume to the Murnaghan equation of 
state \cite{murnaghan}. Then the nanocrystals were passivated 
with hydrogen atoms in the case
of silicon and  with pseudo-hydrogens \cite{passivation-rule} 
in the case of binary nanocrystals.
All atoms of the nanocrystals were relaxed to attain 
the minimum energy configuration. An average bond-length was determined
by averaging over all the nearest-neighbor bond-lengths. This was
then used to determine an average equilibrium lattice constant.
While generating the nanocrystals we can start with either an anion 
or a cation at the center. In the present case we have considered both 
schemes of generation of the binary nanocrystals. 
As the conclusions arrived at were similar in both cases we present the 
results for only one case.

We find it useful to use the definition of Masadeh et. al \cite{billinge-pdf} 
to define the surface stress. The surface stress generated in the 
nanocrystals is defined as

\begin{equation}
Bond strain (\%) = \frac{(r_{0}-r)}{r_{0}}
\end{equation}

where '$r_{0}$', is the nearest neighbor bond-length between the atoms,
is calculated using the theoretical equilibrium lattice constant for the 
bulk. 'r' is the nearest neighbor bond-length that one obtains after 
the optimization of the structure described earlier. It should be noted
that "r" varies from shell to shell of the cluster, and could be 
different even for atoms of the same type in a given shell. This 
arises because of the differences in surface coordination 
that one can have because of the truncation scheme considered. Inspite 
of these variations, the qualitative aspects are similar.

We consider the case of an elemental semiconductor, Si. The calculated 
equilibrium lattice constants are shown in Table II as a function of 
cluster size. One finds that the equilibrium lattice constant is 5.407 $\AA$
for the bulk. The lattice constant computed for the nanocrystals is 
smaller than the bulk in all the cases, as expected from the naive 
considerations presented earlier. A strong size dependence of the 
lattice constant is found in the unpassivated case which shows deviations 
ranging from 2.27 \% to 1.02 \% when we go from smaller clusters with 
3 layers around the central layer to larger clusters with 6 layers around 
the central layer. However the size dependence of the lattice constant 
is much smaller for the passivated nanocrystals, with deviations
$\sim$ 0.3\% - 0.4\% and approaches bulk-like 
values for very small cluster sizes. Similar conclusions are arrived at for
nanocrystals of CdSe and GaAs.

Inspite of the fact that the bulk lattice constant is reached quickly in the 
present case, we do find deviation in the bond-lengths as a function of 
depth from the surface, being maximum at the surface. Quantifying 
this in terms of the bond strain (Fig 1(a)) we find that in the 
core of the nanocrystal, for the size considered we find $\sim$ 0 \% 
bond strain. Beyond the second layer, the strain exhibits a 
linear variation. Since we have a single parameter that changes 
i.e. depth, we would expect a linear variation with depth. 
This is however not the case, and we find that the strain is 
invariant between the third and the sixth layers.
However between the first and third layers one has a linear 
variation with depth, with the core showing almost bulk-like 
lattice constants.  The unusual strain profile that we find here is probably 
a result of a competition between the microscopic considerations 
determining the strain profile in the interior being different 
from those determining it at the surface. Hence the bond strain 
profile depends on the strength of the surface passivant. Consequently, 
the strain profile is different in the case of GaAs (Fig.1(b)) and 
CdSe (Fig.1(c)).

In the case of CdSe we considered both the zinc blende as well as the 
wurtzite polymorphs. As in the zinc blende case, the average lattice constant 
is almost bulk-like for clusters larger than 15 $\AA$ diameter. The bondlength 
profile as a function depth (Fig.1(d)) is however very different. The 
equatorial and the axial bonds show very different depth dependences, showing stronger
anisotropies in the core region than in the bulk, in addition to an 
oscillatory dependence with depth. The latter could arise because the 
wurtzite structure has a finite dipole moment. Interestingly the averaged bond length 
for each layer follows a depth profile very close to one's naive expectations 
being constant upto the third monolayer and then showing a linear 
variation with depth.

We have examined the deviation from bulk-like lattice constants considering 
nanocrystals of Si, GaAs and CdSe. Naive arguments based on broken 
coordination at the surface lead us to expect shorter bonds at the surface 
of the nanocrystal. Indeed this is found to be the case, though the 
surface stress generated is found to be typically $\sim$ 1\% or 
less when we consider nanocrystals passivated by hydrogen or 
pseudo-hydrogens, with almost bulk-like bond-lengths obtained 
5-6 layers below the surface. Averaging over all the bond-lengths 
of the cluster we find that bulk-like bond-lengths are obtained 
for small cluster sizes.

This work has been supported by the Indo French Centre for Promotion
of Advanced Scientific Research.
\renewcommand

\newpage
\begin{table}
\caption
{Atom 'A' centered spherical binary nanocrystals (AB) considered with an underlined
zinc-blende and wurtzite geometry. $N_{A}$ and $N_{B}$ are the number of 'A' and
'B' type of atoms in each nanocrystal.}
\begin{tabular}{|c|c|c|c|c|}
\hline\hline
nanocluster & cluster size (n)  & $N_{A}$ & $N_{B}$ \\ \hline\hline
Zinc-blende &  6 & 79 & 68 \\
 & 5 & 43 & 44 \\
 & 4 & 19 & 16 \\
 & 3 & 13 & 16 \\ \hline
Wurtzite & 5 & 51 & 41 \\
 & 4 & 19 & 20 \\
 & 3 & 13 & 14 \\
\hline\hline
\end{tabular}
\end{table}

\begin{table}
\caption {Comparision of the equilibrium lattice constant and
average bond strain for the unpassivated Si nanocrystals with the 
average equilibrium lattice constant and average bond strain for the passivated
Si nanocrystals as a function of cluster size.}
\begin{tabular}{|c|c|c|c|c|c|c|c}
\hline\hline
cluster size (n)  & diameter ($\AA$) & \multicolumn{2}{c} {lattice constant ($\AA$)} \vline  & \multicolumn{2}{c} {average bond strain (\%) } \vline \\
\cline{3-6}
 & &unpassivated & passivated & unpassivated & passivated \\
\hline\hline
$\infty$ &  $\infty$ & 5.407 & -  & 0.0 & - \\
6  & 17.10 & 5.352 & 5.392 & 1.02 & 0.28 \\
5  & 14.05 & 5.345 & 5.384 & 1.15 & 0.43 \\
4  & 10.81 & 5.323 & 5.379 & 1.55 & 0.52 \\
3  & 8.97  & 5.284 & 5.384 & 2.27 & 0.43 \\ 
\hline\hline
\end{tabular}
\end{table}

\renewcommand
\newpage

\begin{figure}
\includegraphics[width=5.5in,angle=270]{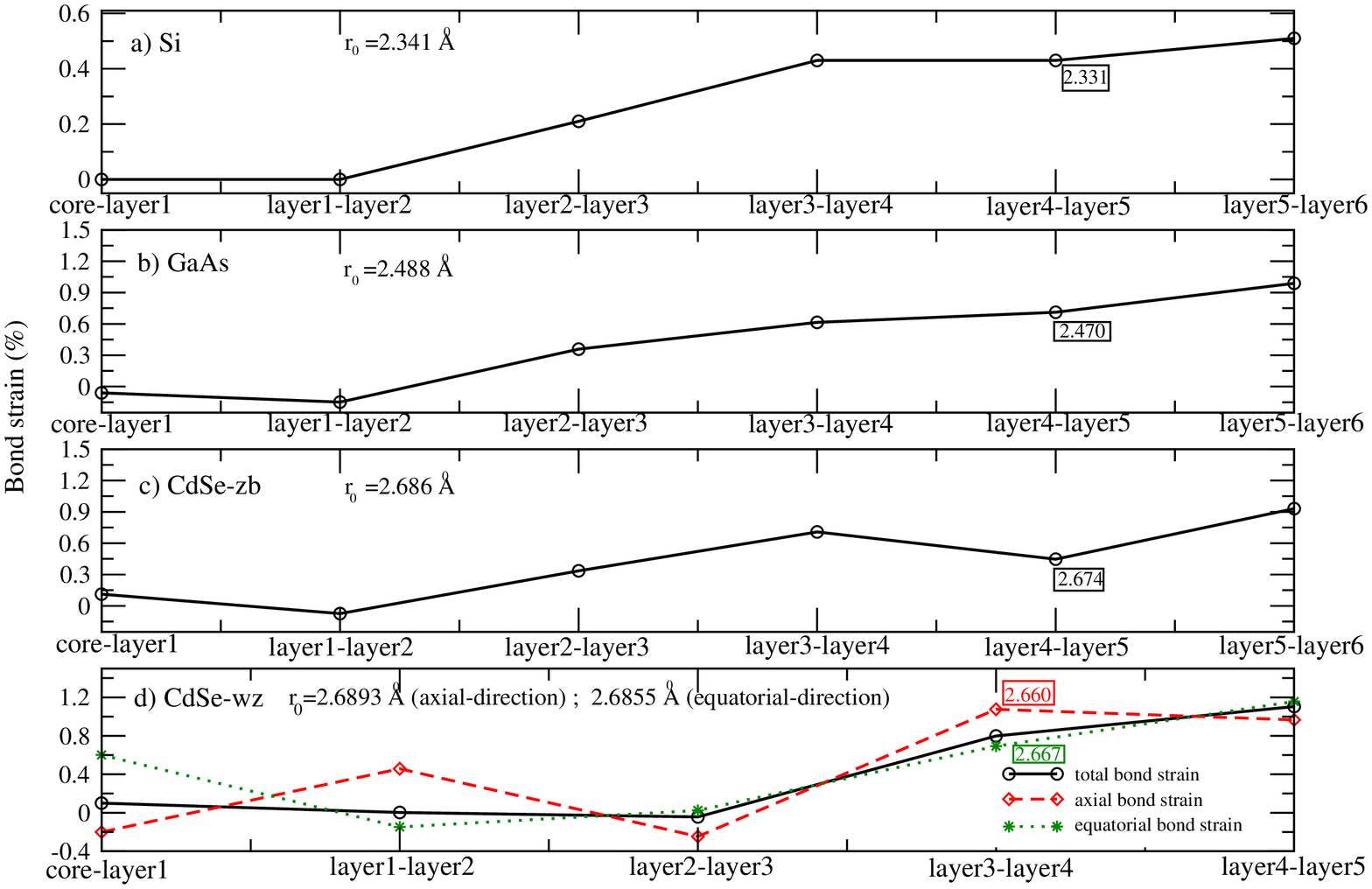}
\caption{Calculated bond strain variation between indicated layers (x-axis) 
for a) Si (n=6), b) Ga-centered GaAs (n=6), (c) Cd-centered zinc blende CdSe (n=6) 
and d) Cd-centered wurtzite CdSe (n=5) nanocrystals.
The corresponding nearest neighbor bond-length between the atoms at the theoretically obtained 
equilibrium bulk lattice constant for each case is given by $r_{0}$. 
In the case of wurtzite nanocrystals the axial (red dashed line), equatorial 
(green dotted line) as well as the average bond strain (black solid line) 
for each layer have been shown as a function of depth.
The values of the bond length for some layers have been indicated. 
A positive strain corresponds to compression of bonds by definition.}
\end{figure}


\newpage

\begin{references}
\bibitem{billinge-science}
S. J. L. Billinge, and I. Levin, Science {\bf 316}, 561 (2007).
\bibitem{heron}
N. Herron, J. C. Calabresse, W. E. Farneth, and Y. Wang, Science {\bf 259}, 1426 (1993).
\bibitem{billinge-pdf}
A. S. Masadeh, E.  Bozin, C. L. Farrow, G. Paglia, P. Juhas, S. J. L Billinge, A. Karkamkar, and 
M. G. Kanatzidis, Phys. Rev. B {\bf 76}, 115413 (2007).
\bibitem{conflict}
As there are conflicting reports of CdSe nanoparticles favoring the zince-blende/wurtzite 
structure, we have considered both polymorphs.
\bibitem{gan-clusters}
G. M. Dalpian, M. L. Tiago, M. L. Puerto, and J. R. Chelikowsky, Nano Lett., {\bf 6}, 501 (2006);
R.Cherian, and P. Mahadevan, Phys. Rev. B {\bf 76}, 075205 (2007).
\bibitem{bulk-like}
M. A. Marcus, L. E. Brus, C. Murray, M. G. Bawendi, A. Prasad, and A. P. Alivisatos, Nanostructured 
Materials {\bf 1}, 323 (1992); P. J. Wu, Y. P. Stetsko, K. D. Tsuei, R. Dronyak, and K. S. Liang, 
Appl. Phys. Lett. {\bf 90}, 161911 (2007).
\bibitem{vasp}
G.~Kresse, and J.~Furthm$\ddot{u}$ller, Phys. Rev. B {\bf 54}, 11169 (1996); G.~Kresse, and J.~Furthm$\ddot{u}$ller, Comput. Mat. Sci. {\bf 6}, 15 (1996).
\bibitem{ggapw91}
J. P. Perdew, and Y. Wang, Phys. Rev. B {\bf 45}, 13244 (1992).
\bibitem{murnaghan}
F. D. Murnaghan, Proc. Natl. Acad. Sci. U.S.A. {\bf 30}, 244 (1944).
\bibitem{passivation-rule}
X. Huang, E. Lindgren, and J. R. Chelikowsky, Phys. Rev. B {\bf 71},165328 (2005).
\end{references}
\end{document}